\documentclass[fleqn,usenatbib]{mnras}

\usepackage{newtxtext}
\usepackage[T1]{fontenc}

\DeclareRobustCommand{\VAN}[3]{#2}
\let\VANthebibliography\thebibliography
\def\thebibliography{\DeclareRobustCommand{\VAN}[3]{##3}\VANthebibliography}

\usepackage{graphicx}	% Including figure files
\usepackage{amsmath}	% Advanced maths commands
\usepackage{amssymb}	% Extra maths symbols
\usepackage{diagbox}
 
\title[NGC~5236's stars as tracers of arms]{NGC~5236's stars as tracers of arms and arm widths in spiral galaxies}

\author[Silva-Villa et al.]{
Silva-Villa, E.,$^{1}$\thanks{E-mail: esteban.silvav@udea.edu.co}
Cano Gómez, X.,$^{1}$
\\
$^{1}$Physics Institute, University of Antioquia, Calle. 67 No. 53-108, A. A. 1226 Medellín, Colombia\\
}

\date{Accepted 2022 April 27. Received 2022 April 26; in original form 2022 March 29}

\pubyear{2021}

\begin{document}
\label{firstpage}
\pagerange{\pageref{firstpage}--\pageref{lastpage}}
\maketitle

\begin{abstract}
Generally, identifying the spiral arms of a spiral galaxy is not a hard task. However, defining the main characteristics, width and length of those structure is not a common task. Previous studies have used different tracers: Star clusters, Massers, H$\alpha$. It was until recently that individual stars were used as tracers of spiral structures. The basic method of measuring the width of spiral arms assumes a Gaussian distribution around the mean concentration, either of gas or other tracer. In this work we use NGC~5236's stars as tracers. We estimated the surface stellar density of arms and inter-arm regions to measure the width of the arms. As a test case, this works focused on NGC~5236 (M83). We find that field stellar populations can trace the (two) main spiral arms of NGC~5236. We find a correlation between the arm width and the galactocentric radii, found using other tracers. The slope of the growth of the width of the arm correlates with the morphological types of spiral galaxies. A second finding of our study suggest the possible correlation between the width of the arms and the corrotation radius, result that will be presented in a follow up paper.
\end{abstract}


%%%%%%%%%%%%%%%%%%%%%%%%%%%%%%%%%%%%%%%%%%%%%%%%%%

%%%%%%%%%%%%%%%%% BODY OF PAPER %%%%%%%%%%%%%%%%%%

\section{Introduction}

In 1926 Hubble created his method of galactic classification based on the morphological characteristics of galaxies (known as the Hubble Sequence). He divided galaxies into three large groups: elliptical (e.g. $E0-7$), spiral (e.g. $Sa-c$) and irregular, plus a transition group between spiral and elliptical galaxies, called lenticular galaxies \citep{Hubble1926}. Later,  \cite{Vaucouleurs1959} added to the Hubble sequence important features of spiral galaxies such as half-bars, rings and lenses. 

Although the Hubble classification only takes into account the morphology of galaxies, this classification is also closely related to dynamical and evolutionary processes of each galaxy type. In particular, the classification for spiral type galaxies may vary according to the shape of the bulge and the number of arms. These galaxies are classified in the Hubble sequence from those with more coiled arms (Sa) to galaxies with more open arms  \citep[Sc, ][]{DobbsBaba2014}. The work by \cite{Vaucouleurs1959} includes in his classification Sd type galaxies, which have poorly defined and irregular arms. Some galaxies have a bar structure in the bulge from which the arms wrap around the galaxy. We now know that galaxy arms are an overdensity of stars, gas, and dust in the galactic disk. Galactic arms are dominant in the visible images because of its high star formation when compared to the inter-arm regions \citep[e.g.][]{SilvaLarsen2012,moss13}.

An unsolved problem in astronomy is to establish a full characteristics of spiral arms in disk galaxies. \cite{Poggio2021}, \cite{Mosenkov2020}, \cite{HonigReid} and \cite{SilvaLarsen2012} are some of the authors who have studied the problem of defining the arms in spiral galaxies. The study done by \cite{Mosenkov2020} made use of integrated light at different wavelengths, while \cite{HonigReid},  \cite{SilvaLarsen2012} among others have made use of observations of gas regions taken in the $H_{\alpha}$ band. \cite{HonigReid} found that the pitch angle cannot be a trust-full parameter to characterize arms in spiral galaxies due to its internal changes at different radii. 

Despite the physical property(ies) used to trace and reconstruct the arms in spiral galaxies, there is no final consensus regarding a particular parameter that can fully describe an arm.

Based on the assumption that arms are defined by the concentration of gas and/or stars, in this work we seek to define the arms of the galaxy NGC~5236 (M83) from the density of individual star as a tracer.  This type of tracer has not been used for the characterization of the arms of disk galaxies, with the exception of our galaxy, which was recently studied by \cite{Poggio2018,Poggio2021}. 

This paper has the following structure: In Sec. \ref{sec:obs} we briefly describe the observations used and the method use to identify the arms of NGC~5236; in Sec. \ref{sec:results} we present our results. Finally, in Sec. \ref{sec:disc_conc} we discuss our results and present the conclusions.

\section{Observations}
\label{sec:obs}
\begin{figure}
\includegraphics[width=\columnwidth]{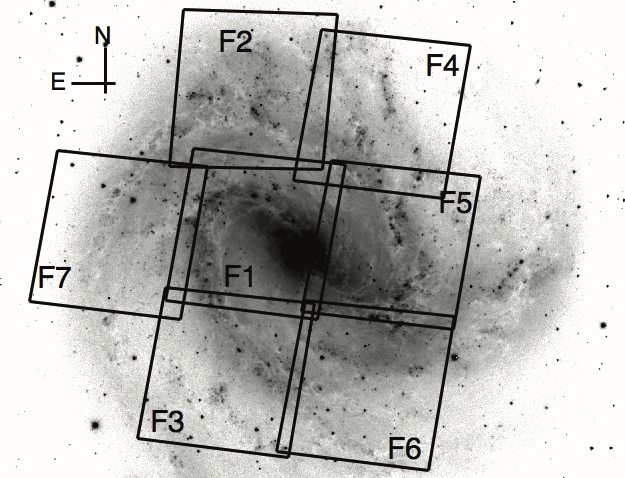}
\caption{R-band image for the galaxy NGC~5236 with overlapping regions observed with HST/WFC3 used in the current work. Credits of the Image: Dr. Angela Adamo.}
\label{fig:mosaic}
\end{figure}

The present study used archival HST/WFC3 images from the program IDs 11360 (PI. O'Connell) and 12513 (PI. Blair). The dataset consists of seven (7) fields observed with the filters F336W, F438W, F547M, F657N, and F814W.  We will use and refer to the F547M/F555W and F814W as V and I, respectively, although no transformations were applied. For one of the fields (Field~1) the F555W filter was used, instead of the F547M filter. A R-band ground-based image (taken from NED) of NGC~5236 with the footprints of the HST mosaic overlayed is presented in Fig.~\ref{fig:mosaic}. The HST zeropoints\footnote{$http://www.stsci.edu/hst/wfc3/phot\_zp\_lbn$} used in this study were: $ZP_U=23.48$, $ZP_B=24.97$, $ZP_V=25.82$ (F555W), $ZP_V=24.74$ (F547M), $ZP_{H_\alpha}=22.31$ and $ZP_I=24.68$.

Previous works have used the star cluster and field stellar populations of this galaxy \citep[e.g.][]{SilvaVilla2014,Ryon2015,Adamo2015}, providing also a quick description of the used procedures. Here, we will describe briefly the procedures and refer the reader to Silva-Villa et all (2022, in prep.) for the full description of the analysis.

NGC~5236's stars were disentangled from extended sources by the use of their {\it color index} (CI: Difference between aperture photometry using 1 and 3 pixels, i.e. $\sim 0.9-2.6$ pc at the distance of this galaxy\footnote{\cite{Thim2003} estimated a distance of $\sim 4.5$ Mpc.}). The CI parameters allows the separation of extended sources from point sources. 

\begin{figure*}
\includegraphics[scale=0.5]{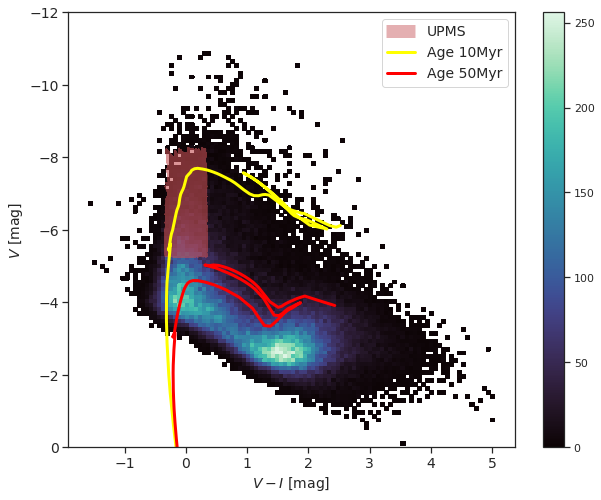}
\includegraphics[scale=0.375]{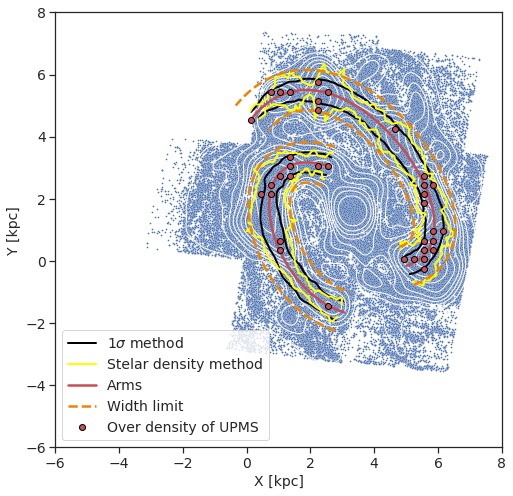}
\caption{{\it Left panel}: Hess diagram of field stellar population in NGC~5236. The yellow and red lines are isochrones of 10Myr and 50Myr, respectively, to identify young stars. The red are are the UPMS used to trace back the arms of the galaxy. {\it Right panel}: Spatial distribution of NGC5236's stars (blue dots) and white contours of over densities. Different lines denote the different methods used in this work, while red dots identify the regions of UPMS stars used to trace the arms. Points identify the over density regions used to trace the arms. See full text for further details. UPMS: upper main sequence stars.}
\label{fig:HR_arms}
\end{figure*}

Stellar photometry was performed over individual fields and standard procedures for point spread function fitting methods were implemented. Figure \ref{fig:HR_arms} (left panel) depicts the color-magnitude diagram (CMD) of our sample. Over the figure, the red area represent upper main sequence stars, defined as the stars with mean color $V-I\approx 0$ within 1$\sigma$ deviation. A similar selection of stars to identify arms in the Milky Way was done by \cite{Poggio2018} and \cite{Poggio2021}.

\subsection{Arms characterization}
\label{sec:arms}

Classification of arms in spiral galaxies generally involves visual inspection of images. Either gas, star clusters, massers or individual stars have been used to determine the regions that define an arm \citep[e.g.][]{SilvaLarsen2012, FShabani2018, Reid2014, Savchenko2020, Poggio2021}. 

We used density of the field stellar population in the galaxy NGC~5236 to identify the location of the arms following the hypothesis that stars form in the highest concentration of gas inside a galaxy \citep[e.g. ][]{Roberts1969}. The red area in the left panel of figure \ref{fig:HR_arms} shows the upper main sequence stars (UPMS), which satisfy the criteria of being massive. We overploted 2 Padova isocrhones \citep{Marigo08} for solar metallicity with ages of $Log(\tau)[yr]=(7,7.7)$, indicating that their age is around a couple of Mys and can be used as tracers. When placed in the $XY$-space, massive bright stars traced the regions where star formation has been recent (i.e. $\tau \leq 10$ Myr, red dots in the right figure). We want to note that we did not estimate ages for the UPMS, we based the selection only on their position in the CMDs. Similar methodologies have been used in the Milky Way \citep{Poggio2018,Poggio2021}. With the identification of the highest density regions of recent star formation, we performed a cubic-spline fit to trace-back the two main arms of the galaxy, as shown in the right panel of the same figure (red lines). We will refer as "arms" the regions determined by the red lines in figure \ref{fig:HR_arms}. The blue dots and the white contours represent the full data set, and it can be seen that the arms estimated follow the highest concentrations of stars. 

The values estimated for tracing back the arms, including the number of stars per box, the galactocentric distance of the box, the width of the arm measured using the 1$\sigma$ method, the width of the arm measured using the density method and their respective errors, a classification of a box with or without a spur (1 and 0 respectively) and a flag to separate Arm1 from the Arm2, can be found in table \ref{tab:table_arms}. The table show only some example of the values, while the full set can be found in a csv file at the GitHub link presented in the Appendix.

Previous studies have measured the widths of arms among different type of spiral galaxies, \citet[e.g.][]{Reid2014, HonigReid, Savchenko2020, Poggio2021}. Previous works have measured widths for spiral galaxies with values reaching 2.5 kpc \citep[e.g.][]{Savchenko2020}. We studied the widths of our initial sample inside that range, i.e. we limited the widths to this value, using a limit value of 1.3 kpc, on each side, perpendicular to the red lines that describe our arms (orange dashed lines in the right panel of Fig.\ref{fig:HR_arms}). 

All previous studies of the width of arms in spiral galaxies based their results on the hypothesis of a Gaussian distribution from the center of the arm outwards (i.e. the mean of the distribution of the tracer used in each study). We studied the distribution following the same procedures but implemented a secondary method based on the surface stellar density. The field stellar density (i.e. the inter-arm region) was estimated using Field 6 from the observations, chosen as the field where the arms showed to be less represented (see Fig. \ref{fig:mosaic}). 

To study the radial variations of the width of the arms, we constructed rectangular regions of size of 1.3 kpc $\times$ 0.3 kpc across the arms, and measured the stellar density inside them. The value measured of each rectangle was studied using histograms of the same bin size. Over each histogram we measured the mean stellar density location and estimated the width of the arm, where the surface stellar density reaches the surface field stellar density as measured from Field 6. We also measured the widths following standard procedures, measuring the 1$\sigma$ of the distribution (assuming a Gaussian profile) in order to compare with previous studies in the literature.

The results of both methods can be found in figure \ref{fig:HR_arms} (right panel, lines yellow and black). 

\section{Results}
\label{sec:results}

Using the density of field stellar populations we traced-back the main (two) arms of the spiral galaxy NGC~5236. Figure \ref{fig:HR_arms} (right panel) present our results for both methods, i.e. using surface stellar density and the standard procedure of measuring 1$\sigma$ from the center of the arm (yellow and black lines respectively).

The values we obtained can be found in a full table following the link in the Appendix. However, table \ref{tab:table_arms} presents the structure of the results obtained.

\begin{table*}
\centering
\begin{tabular}{ |c|c|c|c|c|c|c|c|c| } 
\hline
Box Number & N$_{stars}$ & R$_g$[pc] & Width$_{1\sigma}$[pc] & Error$_{1\sigma}$ & Width$_{SD}$[pc] & Error$_{SD}$& Spur & Arm \\

[1] & [2] & [3] & [4] & [5] & [6] & [7] & [8] & [9] \\
\hline \hline
0 & 1272 & 1628.00 & 291.69 & 5.39 & 443.86 & 0.79 & 1 & 1\\
\hline
1 & 1131 & 1684.44 & 308.10 & 5.44 & 443.42 & 3.51 & 1 & 1\\
\hline
\vdots & \vdots & \vdots & \vdots & \vdots & \vdots & \vdots & \vdots & \vdots \\
\hline
98 & 648 & 4244.42 & 151.64 & 6.68 & 217.16 & 2.73 & 0 & 2\\
\hline
99 & 575 & 4235.01 & 149.18 & 7.08 & 182.27 & 2.79 & 0 & 2\\
\hline
\end{tabular}
\caption{Table with examples of some values used to measure the width of the arms and their relative distance to the center of the galaxy. Column 1 identifies each box used; column 2 depicts the number of stars inside the box; column 3 is the mean galactocentric distance of the box; column 4 is the width of the arm inside the box using the 1$\sigma$ method; column 5 is the error of the width for the 1$\sigma$ method; column 6 is the width of the arm inside the box using the stellar density (SD) method; column 7 is the error of the width for the stellar density (SD) method; column 8 is the flag that state if the box belongs to a spur ($flag=1$) or not ($flag=0$); column 9 is a flag to separate the two arms studied. In this work, only the boxes with stars that are not a spur have been considered.}
\label{tab:table_arms}
\end{table*}

While is true that the possibility of finding spurs/feathers and other possible arms in this galaxy exist, we focused our results in the main two arms that can be observed in optical wavelengths (see Fig. \ref{fig:mosaic}) and selected by eye what could be classified as a spur.

Both methods can trace the regions that follow the two main arms of the galaxy. However, the stellar density method correlate better with the distribution of the arms (see Fig. \ref{fig:HR_arms}, right panel, white contours).

\begin{figure}
\includegraphics[width=\columnwidth]{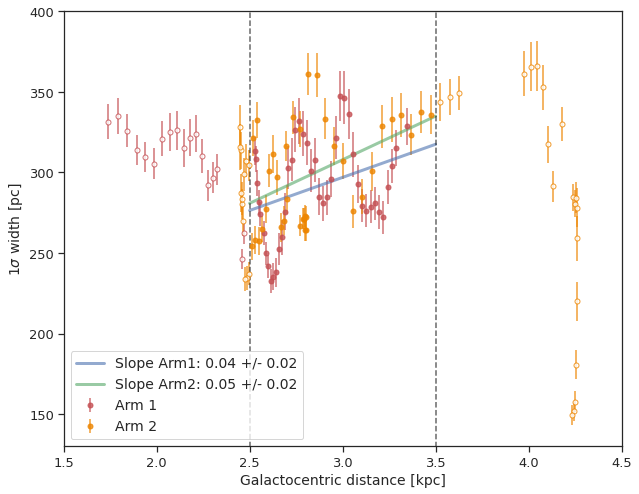}
\caption{Width of the arms at increasing galactocentric distance. The fit follows a straight line with slope as shown in the label and stated in the text. Error bars were estimated using a standard bootstrapping method. The fit is done over the filled symbols of both arms.}
\label{fig:arms_fit}
\end{figure}

% Growing arms width vs galactocentric distance
Based on both methods mentioned before, we estimated the width of the spiral arms of NGC~5236 at different galactocentric radii. We fitted the data in Fig. \ref{fig:arms_fit} using a region where we have values measured for both arms, i.e. $R_{GC}=[2.5,3.5]$ kpc. The fit assumed a linear function (i.e. $deg=1$), which returns the slope, intercept and covariance matrix after minimising the squared errors. The values obtained for the slope of Arm 1 and Arm 2 are $0.04 \pm 0.02$ and $0.05 \pm0.02$ respectively. The errors in the slope are estimated using the value of the corresponding covariance matrix element of the fit scaled by the $\chi^{2}$ divided by the degrees of freedom. The weights are scaled such that the reduced $\chi^2$ is unity\footnote{We refer the reader to the manual of the $Polyfit$ function in Python for further reference.}. For both of the methods used in the study, the widths obtained present a net increase with respect to the galactocentric distance of the arm. We found similar results in the literature using different tracers \citep[e.g.][]{Reid2014,HonigReid,Reid2019}.

The mean of the widths measured in this work for the two main arms of NGC~5236 are: for the 1$\sigma$ method is $0.59$kpc, while for the stellar density method is $0.60$kpc.

\begin{figure*}
\includegraphics[width=\columnwidth]{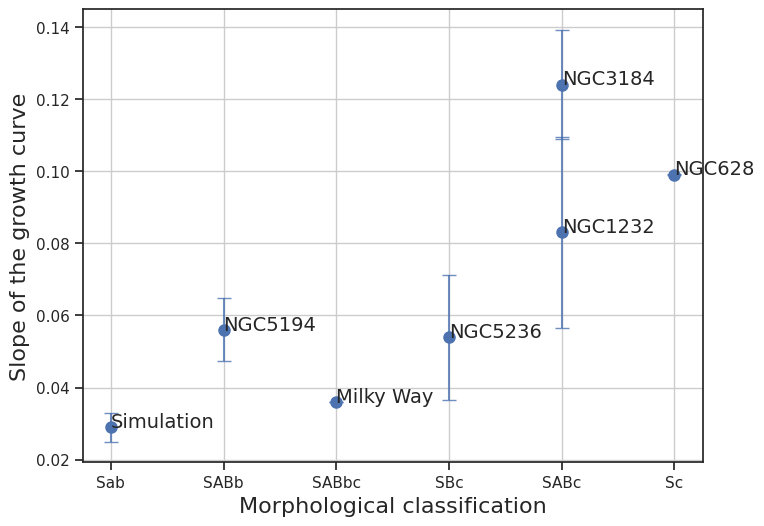}
\caption{Relationship between the morphology and the slope of the growth of the width of the arms of the galaxies for which such information was available \citep[Mikly Way, M51, M74 NGC~1232 and NGC~3184 values were taken from][]{Reid2014,HonigReid}. Error bars were estimated by us using the information in \citet[][see its Fig. 10]{HonigReid} and following the procedure describe in Sec. \ref{sec:results}. The morphological type was taken from the HyperLeda database, except for the simulated galaxy (AM2322-821) for which was used the morphological type reported by \citet{Barros2020} in his undergraduate thesis.}
\label{fig:arms}
\end{figure*}

\section{Discussion and Conclusions}
\label{sec:disc_conc}

% General remarks
Determining the characteristics of spiral arms in galaxies seems to be a not straightforward task. Length, width, pitch angle, can be studied through different tracers, and there is not a common consensus.

To investigate the spiral pattern and star formation activity in the gaseous component of a disk galaxy, \cite{Barros2020} used an N-body simulation with Magnetohydrodynamics (MHD). In his undergraduate thesis he simulated the main component of the AM2322-821 system. The entire simulation runs for 1 Gyr (in code units), but he chose only one snapshot in which the spiral structure is present at $t = 0.3$ Gyr. To obtain the spiral structure he used four different arm tracers: Star Formation Rate, Internal Energy, Free Electrons and Neutral Hydrogen. To measure the width of the arms he used the 1$\sigma$ method. Creating regular cuts at different angles, he constructed a relation between the width of the arm and the galactocentric distance. His results estimated a slope of $0.02 \pm 0.003$ for the correlation.

We compiled the results of previous authors that have estimated the width of spiral arms. \cite{HonigReid} have traced-back the spiral structure of four spiral galaxies using H$II$ regions, including the Milky Way from \cite{Reid2014}. His results are presented in their figure 10, where it can be seen that all galaxies present the same behavior, i.e. the width of the spiral structure grows with galactocentric radii. \cite{FShabani2018} used star clusters as tracers in the galaxy NGC~1566, M51a, and NGC~628 to study the Density Wave Theory predictions, however, the authors did not present any result regarding the growth of the arms. \cite{SilvaLarsen2012} used H$\alpha$ images to trace spiral structure in the galaxy NGC~5236 and estimated its correlation with star formation histories and star formation rate, however, the authors did not study any correlation regarding the width of the arms. \citet{Poggio2018,Poggio2021} study the spiral arms of the Galaxy using the same tracer as we did in this work, however, there are no results regarding the widths of the arms.

The study presented here used surface density of field stellar populations to study the width of the spiral structure in NGC~5236.

% Our results vs other tracers
We found that the surface density of stars can clearly identify the spiral arms of galaxies. Furthermore, we found that there is a positive correlation between the increase in the width of the arms and the galactocentric distance.

% Slope vs morphology
%
\cite{HonigReid} suggested that there is a possible correlation between the width of the spiral arms and the corrotation radius. We found similar results, but we will further investigate this correlation.

% Possible uncertainties
As well as identifying arms in spiral galaxies, its morphological classification can change. Table \ref{tab:table_morphological_types} shows the different morphological classification of the five (5) galaxies shown in this work. In this work we used the classification of Hyperleda.

We study the correlation between the slope of the growth of the width of the arms vs galactocentric distance and the morphological type of galaxies. Combining our result for
NGC~5236 with the measurements of four galaxies by \cite{HonigReid}
, suggests a trend between the slope and the morphological type. However, we warn the reader that, based on the classification used, the results can vary.

One possible way to approach this correlation is using the pitch angle instead of the morphological type. We will investigate this possibility further in a following paper. However, we want to notice that the work by \cite{HonigReid} suggest that the pitch angle for a specify arm can change inside it. This will prevent us to use this parameter as a possible proxy to study the correlation observed.

% Conclusions
There is one only other work where the field stellar population has been used as tracer of the structure of the spiral arms done by \citet[][]{Poggio2021}. However, in their work, the authors did not study the width of the arms. This is the first study that used the surface density of individual stars in a extra-galactic work to measure width of spiral arms. In addition, no previous studies have been found that report the correlation between the growth line of the width of the arms and the morphological type of the galaxies, a relationship shown in Figure \ref{fig:arms}. 

\begin{table}
\centering
\begin{tabular}{ |l|*{5}{c|} } 
\hline
\diagbox{Galaxy }{Catalog} & NED & Simbad & HyperLeda\\
\hline
\hline
Simulation (AM2322-821)& --- & --- & Sab\\
\hline
NGC 5194 (M51) & Sa + Sc & Sc D & SABb\\
\hline
NGC~5236 (M83) & SAB(s)c & SAB(s)c D & SBc\\
\hline
NGC 1232 & SAB(rs)c & SAB(rs)c D & SABc\\
\hline
NGC 3184 & SAB(rs)cd & SA C& SABc\\
\hline
NGC 628 (M74) & 	SA(s)c 
& 	SA(s)c D  & 	Sc\\
\hline
\end{tabular}
\caption{Morphological types of galaxies reported in the databases \href{https://ned.ipac.caltech.edu}{NED}, \href{http://simbad.cds.unistra.fr/simbad/}{Simbad} and \href{http://leda.univ-lyon1.fr}{HyperLeda}. The galaxy AM2322-821 is a simulation, but the authors reported to be an Sab type galaxy.}
\label{tab:table_morphological_types}
\end{table}

\section*{Acknowledgements}

The authors acknowledgement the referee for the comments that helped to improve this work.

%%%%%%%%%%%%%%%%%%%%%%%%%%%%%%%%%%%%%%%%%%%%%%%%%%
\section*{Data Availability}

To retrieve the data used in this work, you can follow the link:

\url{https://github.com/ximenacano1/M83-data}



\bibliographystyle{mnras}
\bibliography{M83-Arms} % if your bibtex file is called example.bib

%%%%%%%%%%%%%%%%%%%%%%%%%%%%%%%%%%%%%%%%%%%%%%%%%%

%%%%%%%%%%%%%%%%% APPENDICES %%%%%%%%%%%%%%%%%%%%%

%\appendix


%If you want to present additional material which would interrupt the flow of the main paper, it can be placed in an Appendix which appears after the list of references.

%%%%%%%%%%%%%%%%%%%%%%%%%%%%%%%%%%%%%%%%%%%%%%%%%%

% Don't change these lines
\bsp	% typesetting comment
\label{lastpage}
\end{document}